\documentclass[12pt,twoside,aps,pra,nofootinbib,amsmath,amssymb,tightenlines,floats,floatfix]{revtex4}
\usepackage{epsfig}
\usepackage{psfig}
\usepackage{mathptmx}
\usepackage{bm}
\usepackage{graphicx}
\renewcommand{\hat}{\widehat}

\begin{document}

\title{Quantum scattering problem without partial-wave analysis}

\author{Vladimir S. Melezhik}

\begin{abstract}
We have suggested a method for treating different quantum few-body
dynamics without usual partial-wave analysis. With this approach
new results were obtained in the physics of ultracold atom-atom
collisions and ionization and excitation/deexitation of helium
ions. The developed computational scheme opens new possibilities
for analysis the actual problem of ultracold atom wave-packets
collisions in strong laser or magnetic confinement.
\end{abstract}

\maketitle

\bigskip

\section{Introduction}
In our papers \cite{Mel91,Mel93} an alternative approach to the
conventional partial-wave analysis was suggested to represent the
few-dimensional Schr\"odinger equation on the basis constructed
from the eigenfunctions of the angular momentum square operator
defined on the angular grid in the spirit of the discrete variable
representation (DVR) \cite{Light} or the Lagrange-mesh method
\cite{Baye}.

To construct the basis orthogonal on the angular grid for one
angular variable is a solvable problem if to chose the grid points
coinciding with the nodes of the Gauss quadrature. Different kinds
of the one-dimensional DVR or Lagrange-meshes are broadly applied
for quantum computations\cite{Light,Baye} due to the simplicity
and efficiency of this approach. However, an extension of this
representation to the two-dimensional case (two angles $\theta$
and $\phi$ of the unit sphere) is a nontrivial problem. Actually,
the simple idea to construct the two-dimensional DVR as a direct
product of two one-dimensional DVRs leads to essential
complication of the matrix of the angular part of the kinetic
energy operator. As a result, the advantages of the
one-dimensional DVR, its simplicity and efficiency, are loosing
\cite{Light}. Another way to construct two-dimensional DVR on an
unit sphere is to use the spherical harmonics defined on the
two-dimensional grid. However, at that it becomes not possible to
satisfy the orthogonality conditions for all the elements of the
fixed set of this basis on the chosen grid. To overcome this
difficulty we have suggested \cite{Mel97,Mel03} to use the basis
of the orthogonalyzed combinations of the spherical harmonics on
the two-dimensional grid over $\theta$ and $\phi$ variables. It
happened that this idea was very efficient for the time-dependent
Schr\"odinger equation with three nonseparable spatial variables
\cite{Mel97}. Particularly, it has permitted to get with this
approach a few important results in problems of the atomic
interaction with strong external electric and magnetic
fields\cite{Mel93,Mel97,Mel00}, the Coulomb breakup of halo nuclei
\cite{Mel99,Mel01,Mel03}.

Recently, we have extended this method to the problem of the
stripping and excitation of helium ions by protons\cite{Mel04} and
antiprotons. We also analysed for the first time the
three-dimensional anisotropic scattering in the problem of
ultracold atom-atom collisions in a laser field
\cite{MelHu,Mesuma}.
\section{Breakup processes and ultracold collisions}

\subsection{Ionization and excitation/deexcitation of helium ions in slow collisions
with antiprotons}

Experimental investigation of the collisions for slow antiprotons
($\overline p$) with hydrogen and helium becomes an actual problem
for antiproton physics. It provides a strong challenge to theory.
Actually, there has been done a large number of theoretical
studies of the $\overline p$-H and $\overline p$-He$^{+}$
recently. However, more or less convergent results were obtained
only for ionization in the collisions $\overline p$-H(1s) and
$\overline p$-He$^{+}$(1s) from the ground states (where there is
an agreement within about 20\% between the existing calculations)
and also some attempts were done for $\overline p$-He$^{+}$(2s)
(see \cite{Sah} and Refs. therein).

We have obtained new theoretical results for ionization and
excitation/deexcitation processes in slow collisions (100 keV
$\geq$ E$_{\overline p} \geq$ 0.1 keV) of ${\overline p}$ with H
and He$^+$ (some of them are given in the Table I). Particularly,
we have calculated for the first time the cross sections from the
initial states with all possible $(l_{i},m_{i})\ne$ 0 to all
possible excitations up to n$_f$=10. The developed quantum
time-dependent approach based on our ideas suggested in
\cite{Mel97,Mel04} opens, thanks to its efficiency and
flexibility, unique possibilities for treating different cascade
processes in antiproton physics.

\begin{table}[h]
\begin{centering}
\caption{Excitation/deexcitation $\sigma_{ex}(n_{i},n_{f})$ and
ionization $\sigma_{ion}(n_{i})$ cross sections from the initial
states $n_{i} l_{i}$ of the He$^+$ for a few antiproton energies
E$_{\overline p}$ (in units of $10^{-18}$ cm$^2$). The cross
sections are summed over final quantum numbers $l_f$ and $m_f$ of
the He$^+$.} \vspace{2ex}
\begin{tabular}{|l|l|l|l|l|l|l|l|l|l|}
\hline \hline ~~~~~~~E$_{\overline p} $ & \multicolumn{3}{|c|}{100
keV} & \multicolumn{3}{|c|}{10 keV} &
\multicolumn{3}{|c|}{1 keV} \\
\hline
 ~~~~~$n_{i} l_{i}$ & 1s & 2s & 2p & 1s & 2s & 2p & 1s
& 2s & 2p \\
\hline $n_{f}=1$ &  & 0.920 & 0.920 &      & 0.667 & 1.07 &      &
8.66 & 8.66\\
$n_{f}=2$ & 8.03 &   &       & 3.99 &      &      & 3.99 &
 & \\
$n_{f}=3$ & 1.59 & 186.0 & 186.0 & 0.927 & 190.0 & 202.0 & 0.927 &
145.0 & 145.0 \\
$n_{f}=4$ & 0.580 & 34.3 & 34.3 & 0.370 & 57.8 & 59.9 & 0.370
& 36.8 & 36.8\\
$n_{f}=5$ & 0.279 & 12.8 & 12.8 & 0.188 & 24.0 & 24.1 & 0.188
& 33.2 & 33.2\\
$n_{f}=6$ & 0.156 & 6.36 & 6.36 & 0.108 & 12.6 & 12.5 & 0.108
& 15.8 & 15.8 \\
$n_{f}=7$ & 0.096 & 3.67 & 3.67 & 0.067 & 7.53 & 7.42 &
0.067 & 9.74 & 9.74\\
$n_{f}=8$ & 0.064 & 2.35 & 2.35 & 0.045 & 4.90 & 4.82 &
0.045 & 6.57 & 6.57 \\
$n_{f}=9$ & 0.044 & 1.58 & 1.58 & 0.031 & 3.34 & 3.28 &
0.031 & 4.65 & 4.65\\
$n_{f}=10$ & 0.032 & 1.12 & 1.12 & 0.023 & 2.40 & 2.35 &
0.023 & 3.45 & 3.45\\
\hline $ionization$ & 9.76 & 93.0 & 93.0 & 4.73 & 180.0 & 162.0 &
4.73 &
86.6 & 86.6\\
\hline \hline
\end{tabular}
\end{centering}
\end{table}

\subsection{Anisotropy effects in control of ultracold atom-atom
collisions by laser fields}

Possible controlling the atom-atom interaction of quantum gases is
an important problem of Bose-Einstein condensation (BEC) at
ultralow temperatures. Applying for that near resonant lasers,
radio frequency fields, Feschbach resonances induced by a magnetic
field, and static electric fields are broadly discussed
\cite{Wei}. We have suggested an alternative possibility: to use a
{\it nonresonant} laser field \cite{MelHu}. Including into
consideration the finiteness of the laser wavelength $\lambda_{L}$
or the alteration of the laser polarization makes the problem of
the atom-atom collisions in the laser field nonseparable over both
angular variables $\theta$ and $\phi$ (scattering angles). It
leads to essentially anisotropic scattering and has demanded to
extend our scheme for a nonseparable three-dimensional stationary
scattering problem \cite{MelHu}.

With this approach we have found considerable influence of a
nonresonant optical laser of intensity $I\geq 10^{5} W/cm^{2}$ on
the Cs-Cs ultracold collisions. In such field the scattering
becomes strongly anisotropic even in the region of ulralow
colliding energies where the $s$-wave dominates at $I=0$. I.e. the
usual scattering length approach $f(k,\hat{\bf k_{i}},\hat{\bf
k_{f}}) = -a_{0}$ does not work and one has to analyze the
stability of BEC for unusual behavior of the amplitude
$f(k,\hat{\bf k_{i}},\hat{\bf k_{f}}) =f(\hat{\bf k_{i}},\hat{\bf
k_{f}})$ at $k\rightarrow 0$. At that the amplitude may be
strongly dependent on the $\lambda_{L}$, on the relative atom-atom
orientation with respect to the field $\hat{\bf k_{i}}$ and on the
scattering angle $\hat{\bf k_{f}}$ \cite{Mesuma}.

The developed computational scheme opens new possibilities for
study the actual problem of few-dimensional wave-packets
collisions in the strong confinement induced by laser or magnetic
traps.

\label{plast}

\begin{thebibliography}{99}
\bibitem{Mel91}
V.S. Melezhik, ``New method for solving multidimensional
scattering problem''{\it J. Comp. Phys. {\bf 92}, 67--81 (1991).}

\bibitem{Mel93}
V.S. Melezhik, ``Three-dimensional hydrogen atom in crossed
magnetic and electric fields'', {\it Phys. Rev. A {\bf 48},
4528--4538(1993).}

\bibitem{Light}
J.C. Light and T. Carrington,Jr., ``Discrete-variable
representations and their utilization'', {\it Adv. Chem. Phys.
{\bf 114}, 263--310 (2000).}

\bibitem{Baye} D. Baye, ``Constant-step Lagrange meshes for central potentials'',
{\it J. Phys. B: At. Mol. Opt. Phys. {\bf 28}, 4399--4412 (1995).}

\bibitem{Mel97}
V.S. Melezhik, ``Polarization of harmonics generated from a
hydrogen atom in a strong laser field'', {\it Phys. Lett. A {\bf
230}, 203--208 (1997).}

\bibitem{Mel00}
{V.S. Melezhik} and P. Schmelcher, ``Quantum energy flow in atomic
ions moving in magnetic fields'', {\it Phys. Rev. Lett. {\bf 84},
1870--1873 (2000).}

\bibitem{Mel99}
{V.S. Melezhik} and D. Baye, ``Nonperturbative time-dependent
approach to breakup of halo nuclei'', {\it Phys. Rev. C {\bf 59},
3232--3239 (1999).}

\bibitem{Mel01}
{V.S. Melezhik} and D. Baye, ``Time-dependent analysis of the
Coulomb breakup method for determining the astrophysical
S-factor'', {\it Phys. Rev. C {\bf 64}, 054612-1-11 (2001).}

\bibitem{Mel03}
P. Capel, D. Baye, and {V.S. Melezhik}, ``Time-dependent analysis
of the breakup of halo nuclei'', {\it Phys. Rev. C {\bf 68},
014612-1-13 (2003).}

\bibitem{Mel04}
{V.S. Melezhik}, J.S. Cohen, and Chi-Yu Hu, ``Stripping and
excitation in collisions between $p$ and $He^{+}$ ($n\leq3$)
calculated by a quantum time-dependent approach with semiclassical
trajectories'', {\it Phys. Rev. A {\bf 69}, 032709-1-13 (2004).}

\bibitem{MelHu}
{V.S. Melezhik} and Chi-Yu Hu, ``Ultracold atom-atom collisions in
a nonresonant laser field'', {\it Phys. Rev. Lett. {\bf 90},
083202-1-4 (2003).}

\bibitem{Mesuma}
V.S. Melezhik, ``Effects of anisotropy in control of ultracold
atom-atom collisions by a light field'', {\it Talk at the Int.
Workshop on Mesoscopic Phenomena in Ultracold Metter: From Single
Atoms to Coherent Ansembles:
http://www.mpipks-dresden.mpg.de/~mesuma04/(Contributions).}

\bibitem{Sah}
S. Sahoo, S.C. Mukherjee and H.R.J. Walters, ``Ionization of
atomic hydrogen and He$^{+}$ by slow antiprotons'', {\it J. Phys.
B: At. Mol. Opt. Phys. {\bf 37}, 3227--3237 (2004).}

\bibitem{Wei}
J. Weiner, V.S. Bagnato, S. Zilio, and P.S. Julienne,
``Experiments and theory in cold and ultracold collisions'', {\it
Rev. Mod. Phys. {\bf 71}, 1--85 (1999).}

\end{thebibliography}
\end{document}